\documentclass[graphics, twocolumn, usenatbib]{mn2e}
\pdfoutput=1
\usepackage{times}
\usepackage{natbib} 
\usepackage{epsfig} 
\usepackage{graphicx} 
\usepackage{color}
\usepackage{aas_macros} 
\usepackage{amssymb}
\usepackage{amsmath}
\usepackage[title]{appendix}
\usepackage{hyperref}	
\hypersetup{colorlinks=true,linkcolor=blue,citecolor=blue,filecolor=blue,urlcolor=blue}
\usepackage[caption=false]{subfig}

\newcommand{\enzo}{\texttt{Enzo~}}
\newcommand{\enzoc}{\texttt{Enzo}}

\newcommand{\kms} {km $\rm{s^{-1}}$}
\newcommand{\mpch} {\rm $h^{-1}$ Mpc\,\,} 
\newcommand{\kpch} {\rm $h^{-1}$ kpc\,\,} 
\newcommand{\msolar} {$\rm{M_{\odot}}~$}
\newcommand{\msolarc} {$\rm{M_{\odot}}$}

\newcommand{\molH} {$\rm{H_2}$~}
\newcommand{\molHc} {$\rm{H_2}$}
\newcommand{\J} {$\rm{10^{-21}\ erg\ cm^{-2}\ s^{-1}\ Hz^{-1}\ sr^{-1}}$}

\newcommand{\smartstar} {\texttt{SmartStar~}}
\newcommand{\smartstarc} {\texttt{SmartStar}}

\begin{document}
\title[]{Fragmentation inside atomic cooling haloes exposed to Lyman-Werner radiation}

\author[J.A. Regan \& T.P. Downes] 
{John A. Regan\thanks{E-mail:john.regan@dcu.ie, Marie Sk\l dowska-Curie Fellow} \& Turlough P. Downes \\ \\
Centre for Astrophysics \& Relativity, School of Mathematical Sciences, Dublin City University, Glasnevin, Ireland\\
\\}

\pubyear{2016}
\label{firstpage}
\pagerange{\pageref{firstpage}--\pageref{lastpage}}

\maketitle

\begin{abstract} 
  Supermassive stars born in pristine environments in the early Universe hold the
  promise of being the seeds for the supermassive black holes observed as high
  redshift quasars shortly after the epoch of reionisation. \molH suppression is
  thought to be crucial in order to negate normal Population III star formation and
  allow high accretion rates to drive the formation of supermassive stars.
  Only in the cases where vigorous fragmentation is avoided
  will a monolithic collapse be successful giving rise to a single massive central object.
  We investigate the number of fragmentation sites formed in collapsing atomic cooling haloes
  subject to various levels of background Lyman-Werner flux. The background Lyman-Werner flux
  manipulates the chemical properties of the gas in the collapsing halo by destroying \molHc.
  We find that only when the collapsing gas cloud shifts from the molecular
  to the atomic cooling regime is the degree of fragmentation suppressed. In our particular case
  we find that this occurs above a critical Lyman-Werner background of J $\sim$ 10 J$_{21}$. The
  important criterion being the transition to the atomic cooling regime rather than the actual value of
  J, which will vary locally. Once the temperature of the gas exceeds T $\gtrsim 10^4$ K and the gas
  transitions to atomic line cooling, then vigorous fragmentation is strongly suppressed. 
\end{abstract}



\section{Introduction} \label{Sec:Introduction}
\begin{table*} 
\centering
\caption{Simulation Parameters}
\begin{tabular}{ | l | c | c | l | c | c | c | c |}
\hline 
\textbf{\em {Sim Name$^{a}$}} &
\textbf{\em {J$_{21}^b$}} & \textbf{\em Maximum Resolution$^{c}$ (pc)} &
\textbf{\em{Collapse Redshift$^{d}$}} &
\textbf{\em{Num Fragments$^{e}$}} 
& \textbf{\em{M$_{\rm{core}}^f$ (\msolarc)}}
& \textbf{\em{M$_{\rm{halo}}^g$ (\msolarc)}}\\
\hline 
Ctrl\_Ref16   & 0.0 & 0.01 & z = 32.68  & 1,1,2  & 2186 & $1.07 \times 10^6$ \\
1J21\_Ref16   & 1.0 & 0.01 & z = 30.55  & 1,3,5  & 6571 & $3.00 \times 10^6$  \\
2J21\_Ref16   & 2.0 & 0.01 & z = 29.91  & 2,5,6  & 5101 & $3.72 \times 10^6$  \\
4J21\_Ref16   & 4.0 & 0.01 & z = 28.87  & 5,6,4  & 9219 & $5.55 \times 10^6$  \\
4J21\_Ref20   & 4.0 & 0.004 & z = 28.87  & 22,-,-  & 12386 &  $5.56 \times 10^6$  \\
4J21\_Ref24   & 4.0 & 0.001 & z = 28.87  & -,-,-  & 12405 & $5.56 \times 10^6$  \\
10J21\_Ref16  & 10.0 & 0.01 & z = 27.15  & 1,1,4  & 12917 & $1.21 \times 10^7$  \\
50J21\_Ref16  & 50.0 & 0.01 & z = 25.36  & 2,4,3  & 15157 & $1.22 \times 10^7$  \\
100J21\_Ref16   & 100.0 & 0.01 & z = 24.73  & 1,1,1  & 16648 & $1.38 \times 10^7$  \\
100J21\_Ref20   & 100.0 & 0.004 & z = 24.72  & 4,-,-  & 20169 & $1.38 \times 10^7$  \\
100J21\_Ref24   & 100.0 & 0.001 & z = 24.72  & -,-,-  & 20359 & $1.38 \times 10^7$  \\

\hline

\end{tabular}
\parbox[t]{0.9\textwidth}{\textit{Notes:} The details of each of the
  realisations used in this study. (a) The simulation name, (b) The LW intensity in units of J$_{21}$,
  (c) the maximum comoving resolution, (d) the collapse redshift (i.e. the redshift at which the
  first smartstar forms), (e) The number of fragments at 80 kyr, 135 kyr and 650 kyr respectively
  (f) mass within the central 1 pc just before the first \smartstar forms
  (i.e. initial mass surrounding the fragmentation site) in solar masses
  and (g) is the halo virial mass in solar masses.
  Note that for the higher resolution runs (i.e. those with 20 and 24 levels
  of refinement) we were not able to follow the evolution for more than 80 kyrs for the 20 levels run
  and only for a few tens of kyrs for the 24 level runs. 
  
}

\label{Table:Sims}
\end{table*}

\noindent Supermassive black holes (SMBHs) populate the centres of massive galaxies
\citep{Kormendy_2013} and shine as extremely luminous quasars out to high
redshift \citep{Fan_2006, Mortlock_2011,Wu_2015}. Their progenitors are unknown, however,
with competing theories for whether the progenitors of SMBHs were stellar mass black holes or
supermassive stars with initial masses of M$_{*,init} \gtrsim 10^4 $ \msolarc
\citep[e.g.][]{Loeb_1994, Madau_2001}. Stellar mass black holes formed from the
remnants of the first stars (PopIII stars) must accrete at the Eddington limit for their
entire history to reach the billion mass threshold by a redshift of 7 if they are to be the seeds
of the first quasars. This scenario appears exceedingly difficult as the first stellar
mass black holes are born ``starving'' \citep{Whalen_2004, Alvarez_2009, Johnson_2011}
in mini haloes which have been disrupted by both the ionising radiation from PopIII stars and the
subsequent supernova explosions \citep{Milosavljevic_2009, Jeon_2014}.
As a direct result, investigation of supermassive star (SMS) formation as a viable alternative has
been undertaken and appears attractive \citep{Haiman_2006, Begelman_2006, Wise_2008a, 
  Regan_2009b, Regan_2009, Volonteri_2010a, Agarwal_2012, Agarwal_2013, Agarwal_2014b,
  Latif_2015b, Latif_2016a, Regan_2017}. The
larger initial seed masses of SMSs alleviate the growth requirements somewhat \citep{Tanaka_2008}.
More fundamentally however, is the fact that the SMS is born in a much larger halo compared to
the canonical PopIII case. SMS formation is thought to be possible only in pristine
atomic cooling haloes with virial temperatures T$_{vir} \gtrsim 10^4$ K with accretion rates exceeding
$\dot{M} \gtrsim 0.1$ \msolarc/yr \citep{Hosokawa_2012, Schleicher_2013, Hosokawa_2013}.
The larger halo leads to a much deeper potential well relative to
the minihalo case, the SMS is also expected to collapse directly into a black hole skipping the
supernovae phase and what's more the radiation spectrum expected from a SMS peaks in the infrared
rather than the UV as is the case for PopIII stars\citep{Hosokawa_2015}.
Taken together, SMSs appear to offer far more favourable conditions for being the original seeds
for SMBHs. \\
\indent The conditions for achieving SMS formation are, however, not yet fully understood. It is widely
suspected that high mass accretion rates combined with high temperature gas will lead to the
formation of a SMS. This can be achieved in a number of ways. Lyman-Werner (LW) radiation from
a nearby galaxy will disrupt \molH abundances removing a coolant, allowing a
halo to grow sufficiently without PopIII formation \citep{Shang_2010, Regan_2014b, Regan_2016a}.
Haloes which accrete unusually rapidly
through a succession of mergers provide another pathway\citep{Yoshida_2003a} as do streaming
velocities\citep{Tanaka_2014, Schauer_2017}
left over from the decoupling of baryons and dark matter after recombination. A combination of
one or more of these scenarios is also possible. We focus here on haloes which are exposed to a
nearby source of LW radiation as it illustrates the necessary points and is straightforward to
implement. The fragmentation of gas inside a collapsing halo could potentially disrupt the
formation of a SMS if the gas were to fragment into sufficiently distant clumps to avoid the
formation of a single (or small multiple) of very massive stars at the centre of a collapsing
atomic cooling halo. \\
\indent We note that any mechanism (with or without a LW component) which suppresses PopIII star
formation and induces the same kind of thermodynamic change to the gas as an external LW field will
also likely impact fragmentation in the same way. While the fragmentation of primordial haloes
has been thoroughly investigated for the case of PopIII
formation\citep[e.g.][]{Bromm_1999, Abel_2000, Clark_2011a, Stacy_2016,
  Safranek-Shrader_2016}
it has not been as systematically probed for the case of irradiated haloes
(but see \cite{Safranek-Shrader_2012}, \cite{Latif_2014b} (L14b), \cite{Regan_2014a} and
\cite{Inayoshi_2014b} for some preliminary work). Here we investigate this scenario
more systematically, varying the intensity of the radiation and investigating the
degree of fragmentation as a function of resolution and radiation intensity. 
\begin{figure*}
  \centering 
  \begin{minipage}{175mm}      \begin{center}
      \centerline{
        \includegraphics[width=18cm]{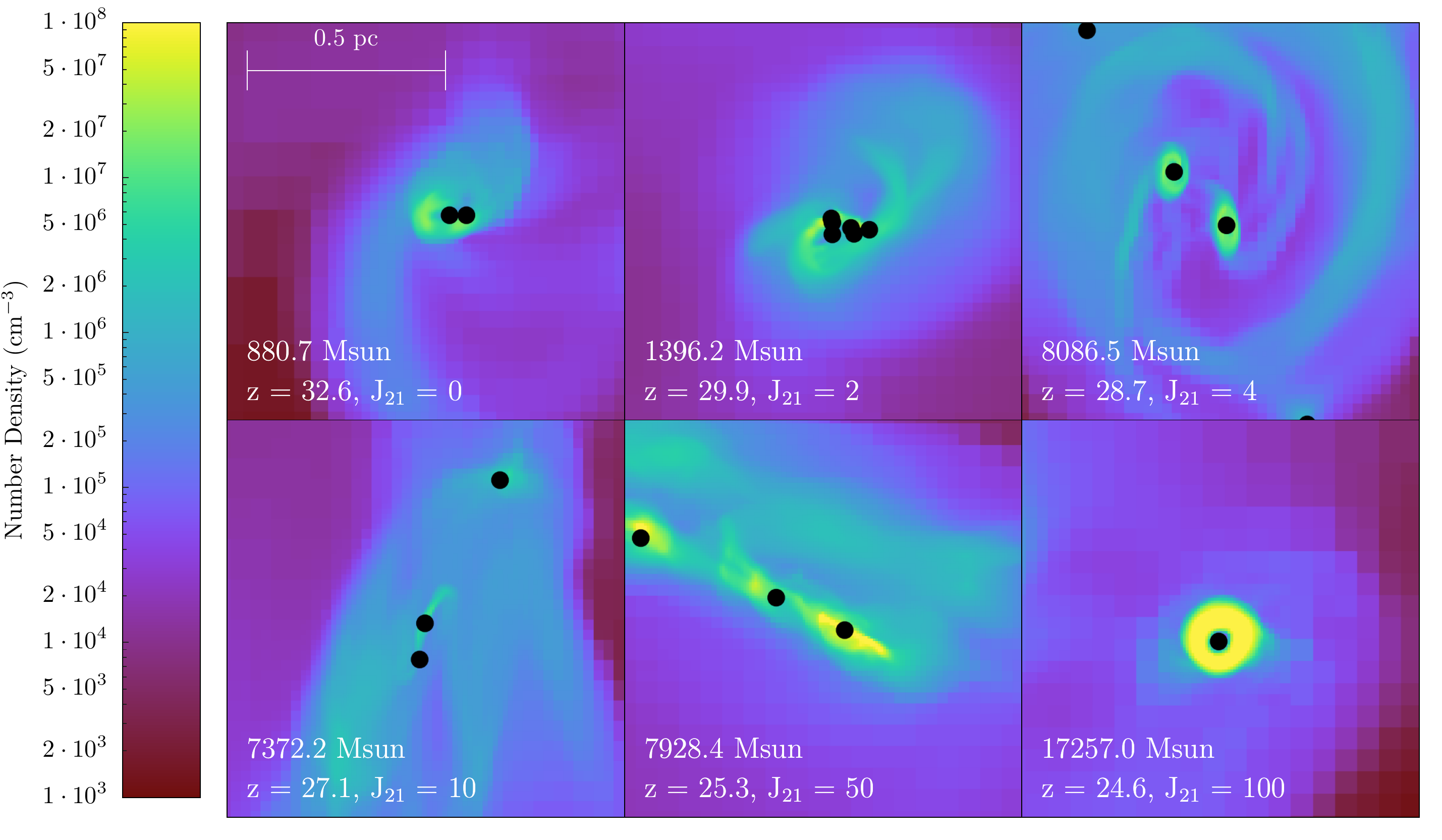}}
        \caption[]
        {\label{Visualisation}
          Projection of the number density through a 1 parsec volume surrounding
          the largest \smartstar particle in each case. The Lyman-Werner background
          radiation produced is varied from panel to panel from 0 J$_{21}$ to 100 J$_{21}$.
          Each realisation is therefore identical except for a different LW background field.
          As the radiation intensity is increased the collapse is delayed to later times
          as \molH is suppressed (note the redshift decreases with increasing LW intensity).
          This raises the Jeans mass of the halo and leads to
          larger \smartstar masses. The projections are created approximately 650 kyrs
          after the formation of the first \smartstar in each case. 
        }
      \end{center} \end{minipage}
  \end{figure*}

The goal of this study is to identify fragmentation sites in collapsing metal-free atomic cooling
haloes which are irradiated by a uniform LW background. As noted above these pristine haloes are
expected to be ideal cradles for SMS formation if the \molH abundances can be suppressed. If
fragmentation levels are reduced in the face of impacting LW radiation, as expected, this would be
favourable for monolithic collapse of a single massive object. We probe this scenario in this work. \\
\indent The paper is laid out as follows: in \S \ref{Sec:Model} we describe the 
model setup and the numerical approach used as well as introducing our star particle formulation;
in \S \ref{Sec:Results} we describe the results of our numerical simulations;
in \S \ref{Sec:Discussion} we discuss the importance of the results and present our conclusions.  \\
Throughout this paper we  assume a standard $\Lambda$CDM cosmology with the following parameters 
\cite[based on the latest Planck data]{Planck_2014}, $\Omega_{\Lambda,0}$  = 0.6817, 
$\Omega_{\rm m,0}$ = 0.3183, $\Omega_{\rm b,0}$ = 0.0463, $\sigma_8$ = 0.8347 and $h$ = 0.6704. 
We further assume a spectral index for the primordial density fluctuations of $n=0.9616$.

\section{Numerical Framework} \label{Sec:Model}
In this study we have used the publicly available adaptive mesh refinement code
\enzoc\footnote{http://enzo-project.org/} to study the fragmentation properties of gas within
haloes irradiated by a background LW field. Into \enzo we have added a new star particle type
which we have dubbed \smartstarc. We now describe both components.
\subsection{Enzo}
\enzoc\footnote{Changeset:fedb30ff370b} \citep{Enzo_2014} is an adaptive mesh refinement code
ideally suited for simulations of the high redshift universe. Gravity in \enzo is solved using
a fast Fourier technique \citep{Hockney_1988} which solves the Poisson equation on the root grid
at each timestep. On subgrids, the boundary
conditions are interpolated to the subgrids and the Poisson equation is then solved at each timestep.
Dark matter is represented using particles, each particle is stored on the highest refinement grid
available to it and thus the particle has the same timestep as the gas on that grid. The
particle densities are interpolated onto the grid and solved at the same time as the gas potential.
\enzo contains several hydrodynamics schemes to solve the Euler equation. We use the piecewise
parabolic method which was originally developed by \cite{Berger_1984} and adapted to cosmological
flows by \cite{Bryan_1995}. The PPM solver is an explicit, higher order accurate version of
Godunov's method for ideal gas dynamics with a spatially third accurate piecewise parabolic
monotonic interpolation scheme employed. A nonlinear Riemann solver is used for shock capturing. The
method is formally second order accurate in space and time and explicitly conserves mass, linear
momentum and energy making the scheme extremely useful for following the collapse of dense
structures. \\
\indent Chemistry is an important component is following the collapse of (ideal) gas. We use the
\texttt{Grackle}\footnote{https://grackle.readthedocs.org/}$^,$\footnote{Changeset:482876c71f73} \citep{Grackle} library to follow the evolution of ten individual species:
${\rm H}, {\rm H}^+, {\rm He}, {\rm He}^+,  {\rm He}^{++}, {\rm e}^-,$ 
$\rm{H_2}, \rm{H_2^+}\, \rm{H^-} \rm{and}\ \rm{HeH^+}$. We adopt here the 26 reaction network
determined by \cite{Glover_2015a} as the most appropriate network for solving the chemical
equations required by gas of primordial composition with no metal pollution and exposed to an external
radiation source. The network includes the most 
up-to-date rates as described in \citet{GloverJappsen_2007, GloverAbel_2008, GloverSavin_2009,
  Coppola_2011, Coppola_2012,  Glover_2015a, Glover_2015b,  Latif_2015}. The cooling mechanisms
included in the model are collisional excitation cooling, collisional ionisation cooling,
recombination cooling, bremsstrahlung and Compton cooling off the CMB.\\
\subsection{Simulation Setup}
\indent \indent All simulations are run within a box of 2 \mpch (comoving), 
  the root grid size is $256^3$ and we employ three levels of nested grids. The grid nesting and
  initial conditions were created using MUSIC \citep{Hahn_2011}. Within the most refined region
  (i.e. level 3) the dark matter particle mass is $\sim$ 103 \msolarc. In order to increase further
  the dark matter resolution of our simulations we split the dark matter particles according to the
  prescription of \cite{Kitsionas_2002} and as described in \cite{Regan_2015}. We split particles
  centered on the position of the final collapse as found from lower resolution simulations within a
  region with a comoving side length of 43.75 h$^{-1}$ kpc. Each particle is split into 13 daughter
  particles resulting in a final high resolution region with a dark matter particle mass of
  $\sim$ 8 \msolarc. The particle splitting is done at a redshift of 40 well before the collapse of
  the target halo. Convergence testing to study the impact of lower dark matter particle masses was
  discussed in \cite{Regan_2015}. \\
  \indent The baryon resolution is set by the size of the grid cells, in the highest resolution region 
  this corresponds to approximately 0.48  \kpch comoving (before adaptive refinement). We vary
  the maximum refinement level (see Table \ref{Table:Sims}) to explore the impact of resolution on
  our results. Refinement is triggered in \enzo  when the refinement criteria are exceeded. The
  refinement criteria used in this work were based on three physical measurements: (1) The dark
  matter particle over-density, (2) The baryon over-density and (3) the Jeans length. The first two
  criteria introduce additional meshes when the over-density
  (${\Delta \rho \over \rho_{\rm{mean}}}$) of a grid cell with respect to the mean density exceeds 8.0
  for baryons and/or DM. Furthermore, we set the \emph{MinimumMassForRefinementExponent} parameter
  to $-0.1$ making the simulation super-Lagrangian and therefore reducing the threshold for
  refinement as higher densities are reached. For the final criteria we set the number of cells
  per Jeans length to be 32 in these runs. \\
  \indent We use 16 levels of refinement as our fiducial refinement level. This corresponds
  to a minimum cell size of $\Delta x \sim 0.01$ pc/h comoving ($\sim 6 \times 10^{-3}$ pc
  at z $\sim$ 30).  We set the effective temperature of the background radiation field to
      T$\rm{_{eff}} = 30000$ K. This background temperature suitably models the spectrum of a
      population of young stars \citep{WolcottGreen_2012, Sugimura_2014, Latif_2015}.
      The effective temperature of the background is important as the radiation temperature
        determines the dominant photo-dissociation reaction set in the irradiated halo. This in turn
        leads to a value of J$_{crit}$ - the flux above which complete isothermal collapse of the
        irradiated halo is observed due to the complete suppression of \molHc. The actual value of
        J$_{crit}$ depends on the nature of the source spectrum \citep{Shang_2010, Sugimura_2014,
          Agarwal_2015a}. \\
        \indent \cite{Agarwal_2015b} proposed that the J$_{crit}$ needed from a given stellar
        population modelled using realistic stellar spectra can vary widely over 2-3 orders of
        magnitude. They argue that in an external pristine atomic cooling halo, DCBH formation is
        better parameterised by using a critical curve in the H$_2$ and H$^-$ photo-destruction rate
        parameter space (further confirmed by \citep{WolcottGreen_2017}). While our choice of
        T$\rm{_{eff}} = 30000$ K falls well within the
        range advocated by these studies, including a realistic source spectrum derived from
        population synthesis models is beyond the scope of the current work and remains to be
        explored in a future study. It should also be noted that \cite{Sugimura_2014} found that
        J$_{crit}$ is only very weakly dependent on the nature of the source spectrum in tension with
        the results of both \cite{Agarwal_2015b} and \citep{WolcottGreen_2017}, however they explored a
        somewhat smaller parameter space.
\begin{figure}
    \includegraphics[width=9cm]{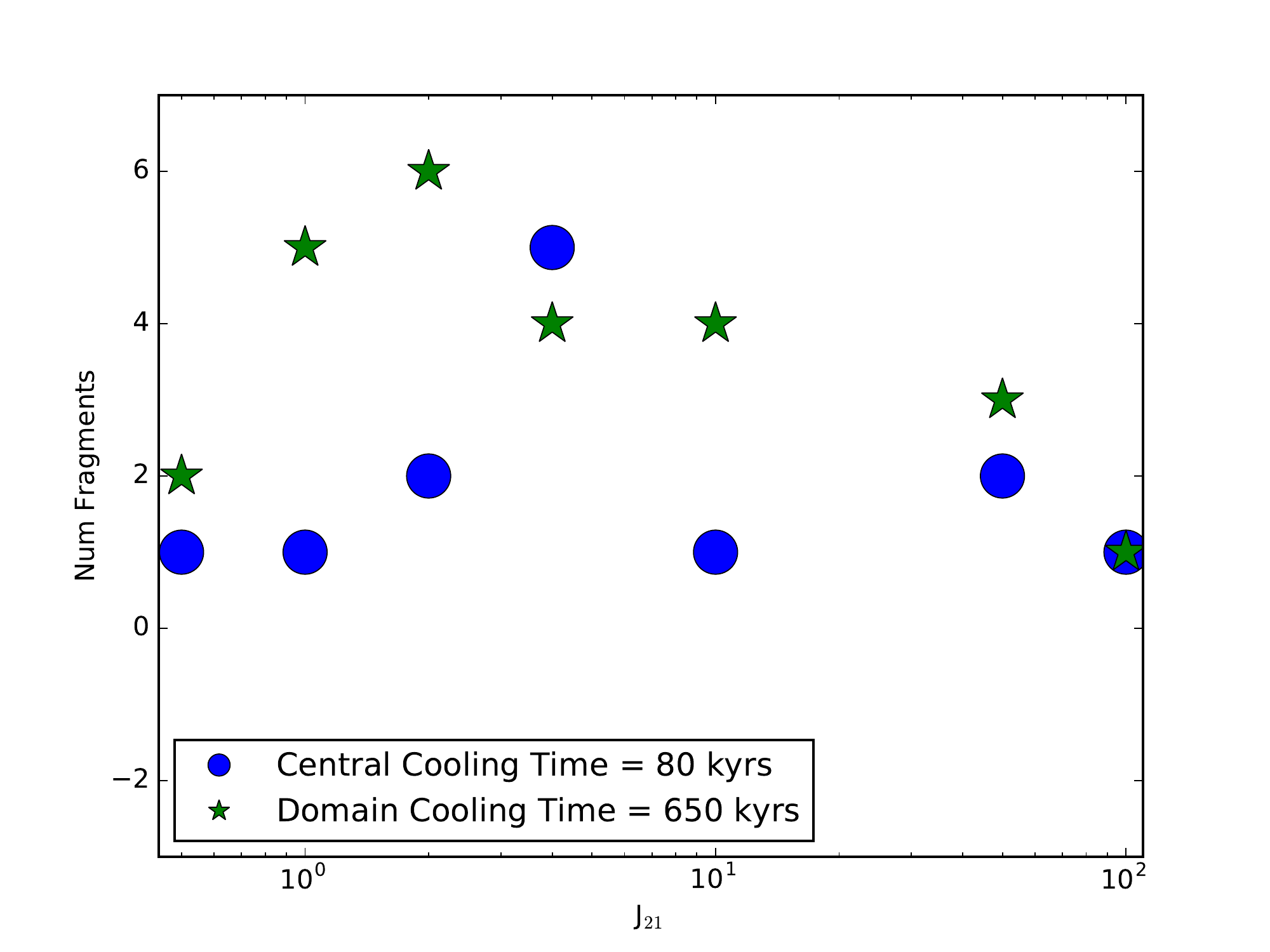}
    \caption[]
            {\label{Fragments}
              The number of fragments which form as the intensity of Lyman-Werner
            radiation is varied. Blue circles represent the number of fragments after
            a central cooling time of 80
            kyrs while green stars represent the number of fragments after
            a domain cooling time of 650 kyrs. The resolution in each case is
            set at $\Delta x \sim 4 \times 10^{-3}$ pc (physical). 
          }
     
\end{figure}

\begin{figure*}
  \centering 
  \begin{minipage}{175mm}      \begin{center}
      \centerline{
        \includegraphics[width=18cm]{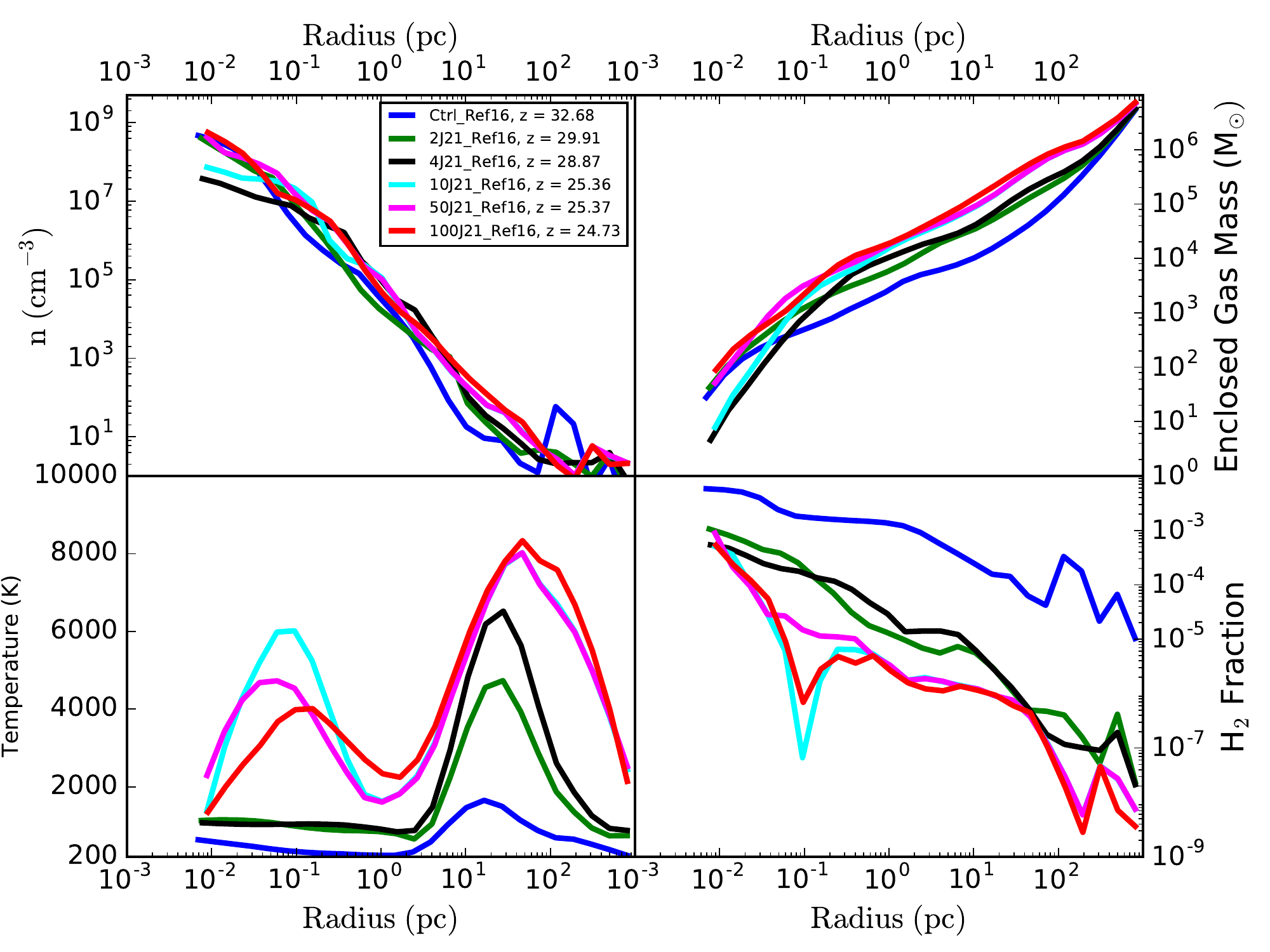}}
        \caption[]
        {\label{RadialProfiles}
          Radial profiles of the gas just prior to the formation of the
          first \smartstar particle. The profiles are for realisations
          with the same maximum resolution (refinement level of 16, 
          see Table 1) but for varying LW intensities. For LW intensities
          of J $\gtrsim$ 10 J$_{21}$ there is a noticeable switch in the temperature
          of the gas (see bottom left panel). For these larger intensities the
          gas is switching to the atomic cooling regime and as we will see
          fragmentation becomes suppressed. 
        }
      \end{center} \end{minipage}
  \end{figure*}


\subsection{Smart Stars}
As the gas density increases in high density regions hydro codes, including \enzoc, require a method
to convert the high density gas into stars in many cases. This is done to
allow gas which has reached the  maximum allowed refinement level of the simulation and for which
further collapse is being artificially suppressed through
artificial pressure support to be dealt with. In this case it can often be prudent
to introduce particles into the calculation to mimic the act of real star formation. It should
however be noted that in many cases there is only a loose correspondence between the numerical
particle and resolving actual star formation. Depending on the level of resolution, primarily, the
numerical particle may represent a single star or an entire cluster of stars. In our study here
the numerical particle introduced will be a proxy for star formation (fragmentation) sites
and represent only regions in which we expect star formation to be likely to occur. Star
(sometimes also known as sink) particles were initially introduced into grid codes by
\cite{Krumholz_2004}. The \smartstar implementation used here also follows this model with some
modifications. We employ the same model for gas accretion as \cite{Krumholz_2004} for our star
particles using a fixed accretion radius of four cells and accreting gas based on the Bondi-Hoyle
prescription. However, while \cite{Krumholz_2004} use only two criteria to evaluate the formation
of sink particles; that the gas is at the highest refinement level and that the cell density exceeds
the Jeans density (see equation 6 in \cite{Krumholz_2004}) we also employ additional sink particle
formation criteria. Similar to the prescriptions described by \cite{Federrath_2010} we
only form sink particles when the following criteria are met:
\begin{enumerate}
\item The cell is at the highest refinement level
\item The cell exceeds the Jeans Density
\item The flow around the cell is converging along each axis
\item The cooling time of the cell is less than the freefall time
\item The cell is at the minimum of the gravitational potential
\end{enumerate}
We calculate the gravitational potential in a region of twice the Jeans length around the cell.
We experimented with also including the additional conditions relating to the gas boundedness and the
Jeans instability test (see \cite{Federrath_2010} for more details). However, we found that these
additional tests were sub-dominant compared to the criteria noted above and so in the interest of
optimisation we did not include them.\\
\indent Once a \smartstar is formed it can accrete gas within its accretion radius (4 cells) and it can
merge with other \smartstar particles. In this study we are focused only on ascertaining the degree
of fragmentation experienced in haloes as a function of Lyman-Werner radiation. The \smartstar
particle type contains algorithms for different accretion modes and feedback modes as well as the
ability to differentiate between normal PopIII star formation and supermassive star formation based
on accretion rates. However, we do not employ these more sophisticated accretion or feedback
algorithms here and instead leave those for an upcoming study. Here as noted previously we focus
only on identifying fragmentation sites in collapsing atomic cooling haloes.
In Table \ref{Table:Sims} we list the simulation parameters and some of the
fundamental \smartstar parameters and results. 

\begin{figure*}
  \centering 
  \begin{minipage}{175mm}      \begin{center}
      \centerline{
        \includegraphics[width=18cm]{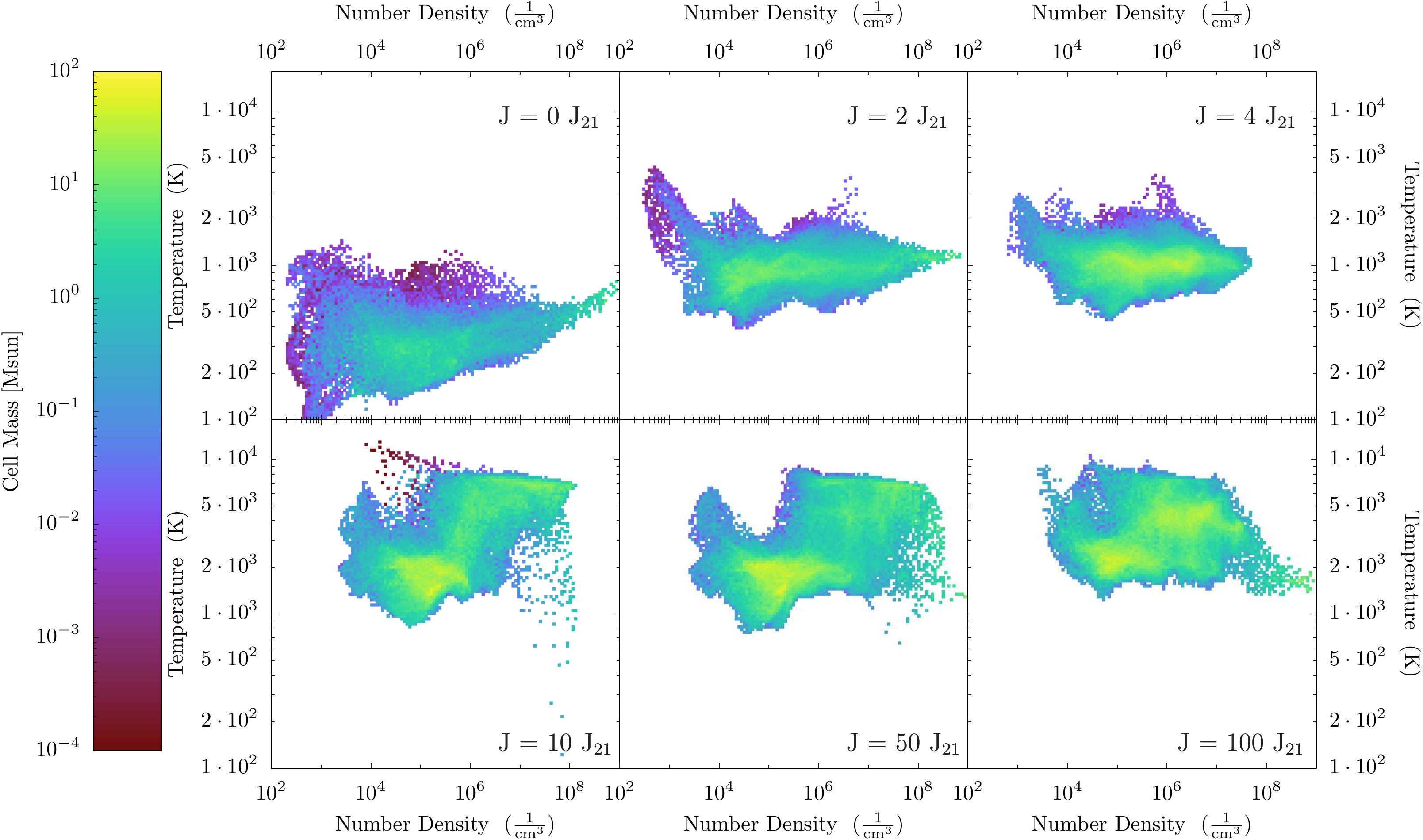}}
        \caption[]
        {\label{Phaseplot_sixpanel}
          The thermodynamic state of the gas as a function of LW intensity
          at fixed resolution (maximum refinement 16). As the LW intensity increases
          the \molH fraction is suppressed altering the cooling function of the gas.
          Initially (J = 0 J$_{21}$) the gas remains cool as the collapse takes hold,
          however, as the intensity increases the gas moves onto the atomic cooling track
          with cooling regulated by atomic cooling at closer to 8000 K.  
        }
      \end{center} \end{minipage}
  \end{figure*}


\begin{figure*}
  \centering 
  \begin{minipage}{175mm}      \begin{center}
      \centerline{
        \includegraphics[width=18cm]{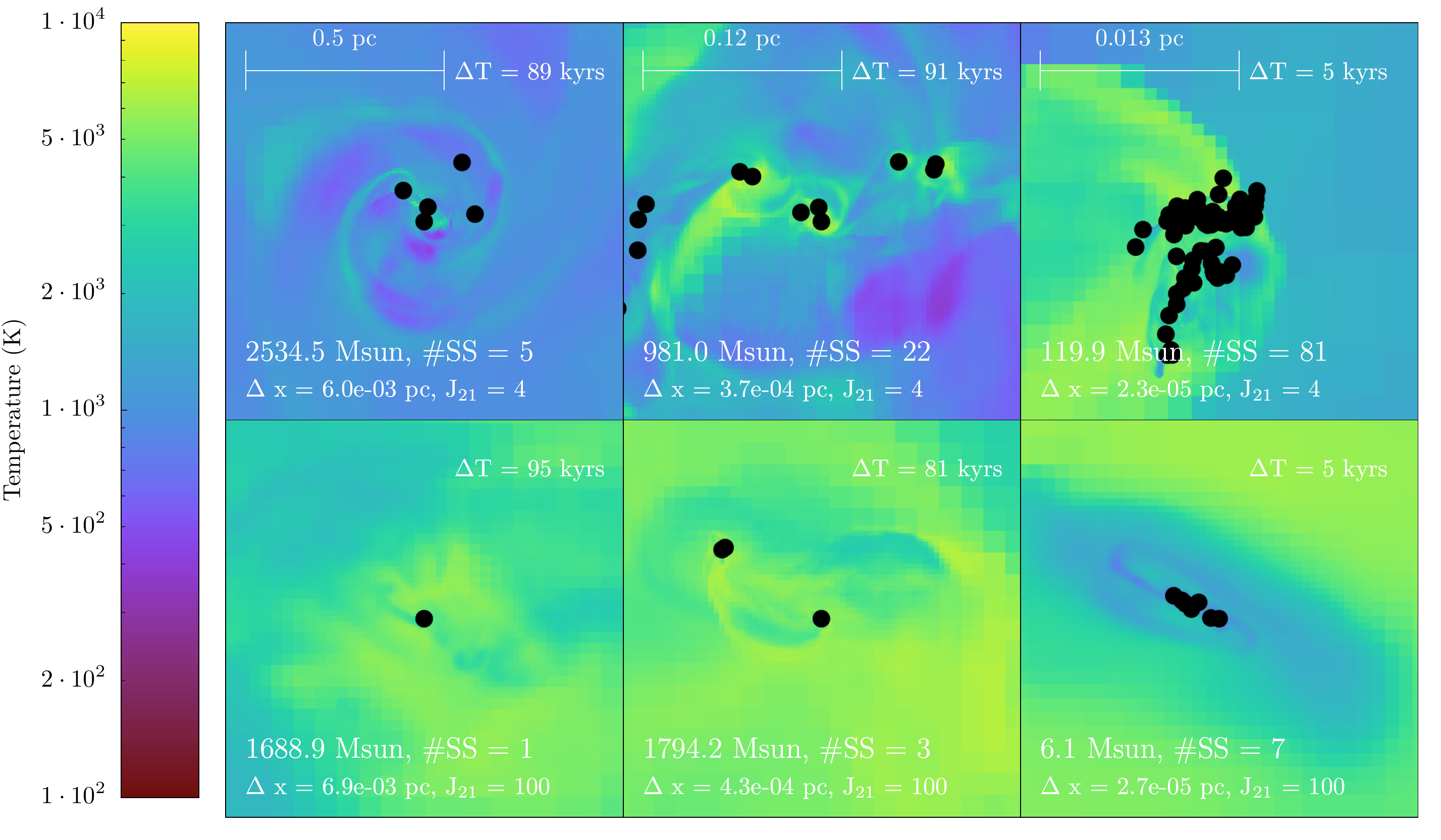}}
        \caption[]
        {\label{Fragments_sixpanel}
          The fragmentation as the resolution is changed for haloes exposed to
          LW intensities of 4 J$_{21}$ and 100 J$_{21}$ respectively.
          The time of the snapshot, $\Delta T$, is given in the top right corner
          of each panel and represents the age of the most massive \smartstar particle.
          Also given is the mass of the most massive particle, the number of
          fragments recorded, the minimum spatial scale resolved and the
          LW intensity. The width of the panel decrease from left to right
          as indicated by the scale line, the scale is decreases so that the individual
          fragments are visible as the resolution is increased. 
        }
      \end{center} \end{minipage}
  \end{figure*}


\begin{figure*}
  \centering 
  \begin{minipage}{175mm}      \begin{center}
      \centerline{
        \includegraphics[width=18cm]{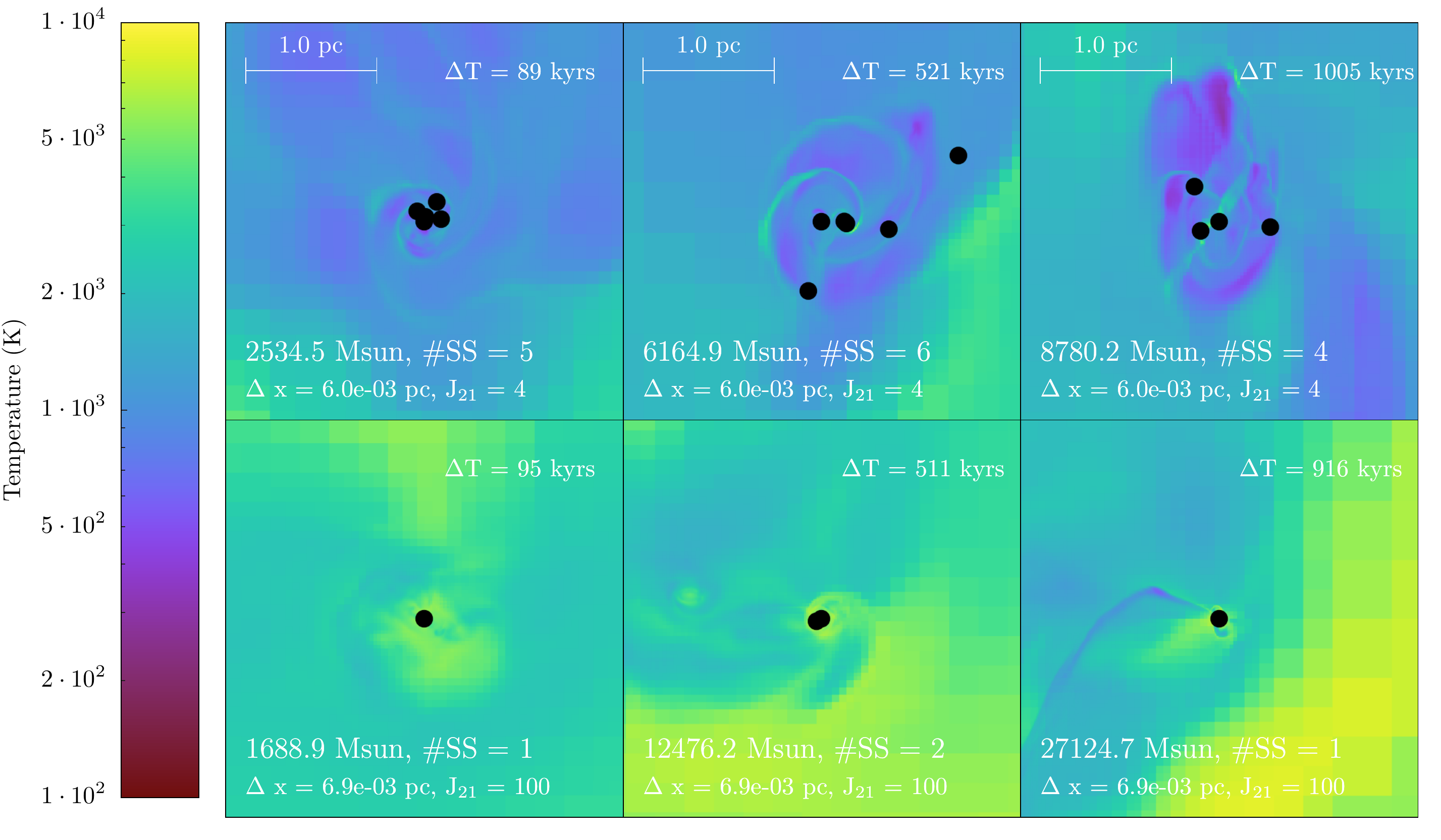}}
        \caption[]
        {\label{Fragments_sixpanel_time}
          The fragmentation for different snapshots for haloes exposed to
          LW intensity of 4 J$_{21}$ and 100 J$_{21}$ respectively. In the left hand
          column the output is approximately 90 kyrs after the formation of the
          first \smartstarc, in the middle column approximately 500 kyrs and the
          right hand column approximately 1 Myr. While some mild fragmentation
          is observed in the 4 J$_{21}$ case it is essentially absent in the
          100 J$_{21}$ case. The time of the snapshot, $\Delta T$, is given in the top
          right corner of each panel and represents the age of the largest \smartstar particle.
          Also given is the mass of the most massive particle, the number of
          fragments recorded, the minimum spatial scale resolved and the
          LW intensity.
        }
      \end{center} \end{minipage}
  \end{figure*}


\section{Results} \label{Sec:Results}

\subsection{The onset of fragmentation as a function of LW intensity}

      In Figure \ref{Visualisation} we show the central 1 pc of the simulations with
      LW intensities of {0 J$_{21}$\footnote{J$_{21}$ is defined as \J},
      2 J$_{21}$, 4 J$_{21}$, 10 J$_{21}$, 50 J$_{21}$, 100 J$_{21}$}. Each panel has an identical
maximum refinement level of 16 levels and hence an equivalent minimum cell size in comoving units.
The projection is made 650 kyrs after the first \smartstar forms. 
Each realisation shows the formation of multiple fragments within the collapsing gas with the
exception of the halo exposed to a radiation level of 100 J$_{21}$. In the final halo of Figure
\ref{Visualisation} (i.e. bottom right panel) only a single
fragment forms at the very centre of the halo and is surrounded by a dense accretion disk
(though mild fragmentation is also observed in this realisation at higher resolution as we will see).
The mass of the fragments show a general trend of increasing mass with increasing radiation level
as expected since the Jeans mass increases in each case due to the suppression of \molHc.\\
\indent In the bottom left of each panel in Figure \ref{Visualisation} we give the mass of the most
massive \smartstar at this time. We do not directly associate this mass with the final mass of the
star that forms since further fragmentation may well, and does (see Figure \ref{Fragments_sixpanel}),
occur below this maximum resolution scale. That is not the goal in this study. Instead we are focused
on identifying regions within the collapsing gas cloud which are likely to experience fragmentation
and likely to foster star formation. Interestingly what we see is that, initially, as the LW
intensity increases, there is a noticeable increase in the number of fragmentation sites. \\
\indent In Figure \ref{Fragments} we have plotted the number of fragments as a function of the
background LW intensity for our fiducial refinement level of 16. We choose two different times
at which to examine the number of fragments. We calculate the cooling time in a given region,
$t_{cool}$, as the average cooling time in a region. The cooling time is the time after
  which the first \smartstar forms that we expect an actual star to form within the fragmentation
  site. The coarser the resolution the longer this time will be. At the other extreme with
  infinite resolution the fragmentation site would be the actual star and
  $t_{cool}$ would be zero. In order to account for our limited resolution we use the
  cooling time as a estimate of the collapse of the gas within the fragmentation site into stars a
  process which occurs well below the grid scale. \\
\indent We initially select each cell and calculate
its Jeans length. We then calculate the average cooling time in a sphere surrounding that cell
with radius its Jeans length. We then loop over all cells in this manner calculating the cooling
over multiple regions to find the cooling time over a given domain. The equation for the domain
cooling time, $\tau_{cool}$, is then given by:
\begin{align}
  t_{cool} &= \ < \sum^{N_{cell}}_{i = 0} t_{cool, i}^{J_v} > \\
  \tau_{cool} &=  {{\sum^{N_{cell}}_{i = 0} t_{cool} \rho_i} \over {\sum^{N_{cell}}_{i = 0} \rho_i}}
\end{align}
where $N_{cell}$ is the number of cells in the domain and $J_V$ is the Jeans volume
  (i.e. the sphere surrounding the cell i of radius the jeans length of that cell) and
  $t_{cool, i}^{J_V}$ is the average cooling time of the gas within that cooling sphere. We then
  sum over all cooling spheres (i.e. all cells) and taking the average cooling time within each sphere
  as the cooling time for that sphere.
We further weight the contribution of each cell in the calculation by it's density with the
highest density cells (those with the shortest cooling times) having a larger contribution. Low
density cells then do not have an appreciable contribution to the calculation. \\
\indent We choose two
different domains. In the first instance we choose the Jeans length of the maximum density cell
and examine all cells within a sphere of radius the Jeans length surrounding that cell
($J_L \sim 0.03$ pc). That leads to an average cooling time of approximately 80 kyrs for the
fiducial runs (i.e. those with a maximum refinement level of 16). The second region we choose
is the entire computational domain. We found through experimentation that the cooling time of
the domain saturates after the region exceeds approximately 10 kpc as outside of this region
the densities become too low and so those cells do not contribute significantly to the computation
of the average cooling time. We find the average cooling time in the total computational domain
corresponds to approximately 650 kyrs. The cooling time will be somewhat dependent on resolution
as well. As higher densities are resolved  we are able to better resolve the actual star
formation timescale and hence our proxy for the star formation timescale (here estimated using the
cooling time) drops. We fix the computation to only calculate the cooling time for the
fiducial refinement level as a benchmark. \\
\indent The first time (80 kyrs) corresponds to roughly when we expect the first actual star to form
after the \smartstar formation has been triggered. This will also correspond to the time at
which we expect radiation feedback effects to begin to impact subsequent fragmentation. The second
time is the other extreme and is the maximum time it could take the surrounding gas to cool and
initiate widespread fragmentation. These two timescales should then bracket the actual fragmentation
timescale - which is unknown and will depend on specific environmental conditions
which are either not resolved by our simulation or not included. In determining the fragmentation
  scale the Jeans length of the gas must be adequately resolved. The Jeans length
  (and cooling time/length) depend sensitively on the temperature of the
  gas. In the 0 J$_{21}$ case the Jeans length is of order $J_L \sim 0.007$ pc at the centre of the collapse
  while it increases to approximately $J_L \sim 0.04$ pc for the cases with radiation. For our
  fiducial runs, $\Delta x \sim 0.004$ pc, and so we are not well resolving fragmentation
  sites for the 0 J$_{21}$ case. The number of fragmentation sites for the 0 J$_{21}$ case can
  only be therefore taken as lower limits. For the cases with radiation, which is the goal of this study,
  we are resolving the Jeans length reasonably well. Nonetheless we include the 0 J$_{21}$ case
  for the case of completeness.\\
\indent For the case of
J = 4 J$_{21}$ the number of fragmentation sites is 5 after 80 kyrs. It is only when the
background is sufficiently strong that it can effect the detailed chemistry of the gas - we then see a
reduction in the number of fragmentation sites. For the highest intensity case probed here
(J = 100 J$_{21}$) the number of fragmentation sites is one both after 80 kyrs and after 650 kyrs.
If we focus on the trend from 2 J$_{21}$ to 100 J$_{21}$ we see the number of fragmentation
  sites decrease as a function of increasing radiation. While there is also a drop from
  1  J$_{21}$ to 0 J$_{21}$ this is likely due to a lack of resolution in the fiducial case and the
  fragmentation levels for the 0 J$_{21}$ are best treated as a lower limit.\\
\indent In Figure \ref{RadialProfiles} we plot the radial profiles of the gas as a function of radius
for the realisation at the same refinement level (Ref16) and varying background intensities. The
plots are made just before the first \smartstar forms in each realisation. The densities reached
are approximately the same in each case since the maximum refinement level is identical and the
densities are close to those at which \smartstar formation is triggered. Note that the times
(redshifts) are different in each case as the LW background delays the onset of cooling within the
halo. Note also how the enclosed mass, \molH fraction and temperature vary according to the LW
intensity. For the highest LW backgrounds the temperature is clearly higher in the central parts of
the halo. These thermodynamic differences effect the fragmentation rates as we saw in
Figure \ref{Fragments}. To further emphasise the temperature and associated thermodynamic
differences we show the phase profiles for each realisation in Figure \ref{Phaseplot_sixpanel}.
As the LW intensity increases we clearly see the state of the gas change reflecting the chemical
changes. As the intensity increases up to and above $10 J_{21}$ we see the gas transition from the
molecular to the atomic cooling regime with a significant mass fraction of the gas above
approximately 2000 K. This transition and the thermodynamic consequences of it are reflected in
the decrease in the level of fragmentation we observe. The increased temperature of a significant
fraction of the gas results in a larger Jeans mass and less fragmentation for the most highly
irradiated haloes.

\subsection{The dependence of fragmentation on resolution}
\indent In Figure \ref{Fragments_sixpanel} we plot the fragmentation as a function of resolution for
two different values of LW intensity - 4 J$_{21}$ and 100 J$_{21}$. In the left hand column the
maximum refinement level is set to 16 ($\Delta x \sim 6 \times 10^{-3}$ pc), in the middle column
the maximum refinement level is increased to 20 ($\Delta x \sim 4 \times 10^{-4}$ pc) and in the
right hand column the maximum refinement level is increased to 24 ($\Delta x \sim 3 \times 10^{-5}$
pc). For the left and middle column the snapshots are plotted at approximately 80 kyrs (i.e. the
cooling time with a Jeans length of the gas for the Ref16 runs) while for the right hand column the
short dynamical times make extending the runs to those times prohibitive and instead we plot them
at 5 kyrs instead (cf. \cite{Stacy_2016}). \\
\indent The projection is made using the temperature field to illustrate the changing chemical
conditions as a function of LW intensity. Note that the spatial scale varies between columns from
left to right. We centre on the most massive \smartstar in each case and as the resolution increases
we are probing fragmentation deeper and deeper inside individual fragmentation sites. As the
resolution is increased we see that the 4 J$_{21}$ realisation shows increased fragmentation, with
the colder gas and subsequently smaller Jeans masses likely facilitating the fragmentation. This is in
contrast with the bottom row where the 100 J$_{21}$ intensities show comparatively little
fragmentation for similar scales. The right hand column shows the most significant fragmentation
demonstrating the most complex structures exist at the smallest scales. Again, however, the 100
J$_{21}$ intensities show over an order of magnitude less fragmentation compared
to the 4 J$_{21}$ intensities at comparable times. The prohibitively short Courant times associated
with the highest
resolution runs (even allowing for the fact that a portion of the high density gas is
tied up in particles) prevents us from evolving the highest resolution simulations for a large number
of dynamical times. However, what is clear is that the level of fragmentation is greatly reduced at
the smallest scales as the LW intensity is increased. 

\subsection{Further evolution of fragmentation}
Finally in Figure \ref{Fragments_sixpanel_time} we explore the time evolution of the realisations
with LW intensity of 4 J$_{21}$ and 100 J$_{21}$ respectively. It should be noted here again that we
do not include the effects of radiation or accretion feedback in these simulations and therefore we
are neglecting some important physical processes which will effect the levels of fragmentation as
star formation proceeds. Additionally, the formation of one \smartstarc, may impact
  physically, on the formation of a subsequent \smartstarc. This would be particularly important
  if the \smartstar releases ionising radiation. For the case of SMS formation with accretion rates
  around 0.1 \msolar yr$^{-1}$ the bloated stellar radii of SMSs lead to effective surface
  temperatures of approximately 6000 K. With these, relatively low, effective temperatures the
  radiation is emitted predominantly in the infrared meaning that it's effect may be quite mild and
  not impact surrounding fragmentation sites. Therefore, for the high external flux cases where
  SMS formation is expected, the formation of one \smartstar is not expected to negatively
  (or positively) impact subsequent SMS formation. If, however, smaller POPIII stars form
  instead then the ionisation front may become important. However, a recent study by
  \cite{Chon_2017b} shows that the ionisation from around POPIII stars which form in close
  proximity to SMSs have small HII regions which are closely bound to the star and do not have
  significant negative impacts at least in the first 100,000 years. 
However, since we do not take these feedback effects into account here, we simply outline the evolution of fragmentation in the absence of
these effects. In Figure \ref{Fragments_sixpanel_time} we see that initially (left hand column) the
\smartstar particles are quite clustered (there is only particle for the 100 J$_{21}$ case). As the
simulation proceeds up to 1 Myr (after which the first \smartstar particle would be expected to go
supernova) the \smartstar fragmentation sites in the 4 J$_{21}$ spread out and lie approximately
1 pc apart. For the 100 J$_{21}$ case the collapse remains completely monolithic. We can also see
that for the 4 J$_{21}$ case the surrounding gas remains comparatively cold compared to the 100
J$_{21}$ case, the temperature of the gas in the 4 J$_{21}$ case is likely to stimulate fragmentation
and hence the increased fragmentation in the lower LW intensity case is not surprising. The main
finding is that the transition to the atomic cooling regime in the 100 J$_{21}$ case leads to and
sustains complete monolithic collapse.

\subsection{Comparison to previous studies in the literature}
\indent L14b and \cite{Latif_2015b} (L15b) performed similar studies to
  that performed here. In L14b their goal was to identify the mass scale of stars formed
  when exposed to a moderate LW background and in L15b they investigated the
  accretion rates on sink particles in the case of non-isothermal collapse. We will focus our
  comparison with the first work - L14b - as it most closely corresponds to our study.
  In L14b they investigated the flux impinging on two separate haloes with LW
  backgrounds of 10 J$_{21}$,
  100 J$_{21}$ and 500 J$_{21}$. Unlike the study performed here they did not systematically
  vary their resolution and did not explore lower LW background cases (i.e. below 10 J$_{21}$).
  Their sink particle prescription was based on that of \cite{Wang_2010} which is broadly similar
  to the prescription we employed although their accretion radius was significantly larger than
  ours ($\sim32$ cells compared to our 4 cells). Nonetheless our results are broadly consistent
  for the relatively high flux cases of J =  10 J$_{21}$ and 100 J$_{21}$ where our two studies overlap.
  The main differences being that we form more sinks for the very high resolution tests
  ($\delta_x \sim$ 1 AU) but with smaller masses. 
  L14b find that one to two sinks form in their very high resolution cases. We find,
  at comparable resolution to their study, between 3 and 7 sinks have formed. Our higher values
  are due to the fact that we have a much smaller accretion radius and hence less merging
  (since the merging radius is identical to the accretion radius in both implementations). Our
  mass scales therefore also differ somewhat, L14b have masses of 7337 \msolar and 2592 \msolar
  for their haloes, for the J = 100 J$_{21}$ after 10 kyr. In our work we see sink particles
  with maximum masses of 1794 \msolar (after 80 kyr) and 6 \msolar (after 5 kyr). Our
  decreased masses can be attributed to our significantly smaller accretion radius.
  Choosing the correct accretion radius is non-trivial and different prescriptions for the
  accretion/merging radius will lead to different results as noted recently by \cite{Becerra_2017}
  in a similar context. Nonetheless, the studies
  give broadly the same result with differences attributable to slightly different subgrid
  implementations. Since their study did not investigate the lower flux cases the degree of
  fragmentation suppression is not available from their study.\\
  \indent Comparing to the results of \cite{Stacy_2016}, which is for the zero flux case,we achieve
  similar resolution and similar results to their very high resolution runs. While they investigate
  the build up of a PopIII IMF we focus on simply identifying fragmentation sites rather than
  individual star formation sites. Encouragingly their fragmentation levels are similar to ours
  (comparing our 4 J$_{21}$ case to their fiducial (0 J$_{21}$) case). After approximately 5 kyrs
  the ionisation front from the first star to form has slowed the accretion rates and subsequent
  star formation in their simulations. They find that of the order of 100 sink particles have formed
  after 5 kyr compared to our 81 sink particles. It is encouraging to see that given our
    resolution levels are comparable that we achieve similar fragmentation levels for the low flux
    cases when the dominant chemistry is still due to \molHc. \\

\section{Summary \& Discussion}  \label{Sec:Discussion}
We investigated here the degree of fragmentation of atomic cooling haloes as a function of both the
strength of the LW background and as a function of resolution. The degree of fragmentation within
irradiated atomic cooling haloes is an important consideration in understanding the zero age main
sequence masses of supermassive stars \citep[e.g.][]{Hosokawa_2013, Woods_2017, Haemmerle_2017}
that are candidates for the supermassive black holes that exist as high redshift
quasars \citep{Haiman_2006, Regan_2009b, Volonteri_2010a}. \\
\indent Haloes that are exposed to high Lyman Werner backgrounds have delayed collapse times and
larger Jeans masses enabling them to form larger objects. However, if fragmentation of the
gas is prevalent the final masses of the (super)massive stars is likely to be greatly reduced. The
thermodynamic conditions of the gas is therefore critical in estimating the initial and
(by accounting for mass loss) the final masses of these objects.\\
\indent We find that haloes exposed to a LW background of J $\gtrsim 10 \rm{J}_{21}$ experience
progressively less fragmentation than haloes exposed to lower LW backgrounds. Haloes exposed to a
LW background of J $\gtrsim 10 \rm{J}_{21}$ have a significant fraction of their gas pushed onto the
atomic cooling track (see Figure \ref{Phaseplot_sixpanel}) and are much less prone to fragmentation.
The transition to the atomic cooling track is likely then to be the most stringent
indicator of fragmentation levels. The increased temperature of the gas in the centre
limits the degree of fragmentation significantly and for the highest irradiation levels examined
here fragmentation was eradicated almost entirely. \\
\indent Comparing to the results of \cite{Stacy_2016} we achieve similar resolution and similar
results to their very high resolution runs. While they investigate the build up of a PopIII IMF
we focus on simply identifying fragmentation sites rather than individual star formation sites.
Encouragingly their fragmentation levels are similar to ours (comparing our 4 J$_{21}$ case to
their fiducial (0 J$_{21}$) case). After approximately 5 kyrs the ionisation front from the
first star to form has slowed the accretion rates and subsequent star formation in their
simulations. Given our resolution levels are comparable it suggests that our assertion that
once the gas chemistry switches to the atomic cooling regime fragmentation is likely to be
strongly suppressed which is something they did not investigate. \\
\indent We note also that any mechanism which suppresses \molH formation and forces a transition
to the atomic cooling track is likely to achieve a similar outcome. For example in the simulations
conducted by \cite{Schauer_2017}, collapsing haloes are exposed to streaming velocities of up to
18 \kms at z = 200. The streaming velocities act in a similar way to a LW background by suppressing
\molH production and hence impacting the gas cooling. The thermodynamic state of the gas is therefore
effected in a similar way to a LW background and hence for haloes exposed to large streaming
velocities fragmentation is also likely to be suppressed. Similarly, dynamical heating effects
\citep[e.g.][]{Yoshida_2003a} combined with mild LW radiation from nearby galaxies as recently
identified by Wise et al. (2017)(in prep) would likely not be effected by vigorous fragmentation
and would in fact be ideal candidates for a single monolithic collapse.\\
\indent In summary we find that haloes exposed to LW radiation experience both delayed collapse
\textit{and} suffer from decreased fragmentation. Once the gas in the core exceeds approximately
2000 K, and has transitioned chemically to predominantly atomic line cooling, then fragmentation
is almost entirely suppressed and conditions becomes ideal for monolithic collapse.
In future work we will explore the impact of feedback from the collapsing
protostar and the subsequent transition to a (massive) black hole seed.

\section*{Acknowledgements}
J.A.R and T.P.D thank David Hubber, Peter Johansson and Bhaskar Agarwal for useful discussion
on both the sink particle implementation and direction of the study. 
J.A.R. acknowledges the support of the EU Commission through the
Marie Sk\l{}odowska-Curie Grant - ``SMARTSTARS" - grant number 699941.
Computations described in this work were performed using the 
publicly-available \enzo code (http://enzo-project.org), which is the product of a collaborative 
effort of many independent scientists from numerous institutions around the world.  Their 
commitment to open science has helped make this work possible. The freely available astrophysical 
analysis code YT \citep{YT} was used to construct numerous plots within this paper. The authors 
would like to extend their gratitude to Matt Turk et al. for an excellent software package.
J.A.R. would like to thank Lydia Heck and all of the support staff involved with Durham's COSMA4
and DiRAC's COSMA5 systems for their technical support. This work was supported by the Science
and Technology Facilities Council (grant numbers ST/L00075X/1 and RF040365). This work used the
DiRAC Data Centric system at Durham University,  operated  by  the  Institute  for  Computational
Cosmology on behalf of the STFC DiRAC HPC Facility  (www.dirac.ac.uk). This equipment was funded
by BIS National E-infrastructure capital grant ST/K00042X/1, STFC capital grant ST/H008519/1,
and STFC DiRAC Operations grant ST/K003267/1 and Durham University.  DiRAC is part of the
National E-Infrastructure. The authors also wish to acknowledge the SFI/HEA Irish Centre for
High-End Computing (ICHEC) for the provision of computational facilities and support. Finally, we
thank the anonymous referee for comments which improved both the quality and clarity of the
manuscript. 

\noindent


\bsp	

\appendix
\section{Smartstar Tests}
As part of the integration of our \smartstar particle into \enzo we rigorously tested the sink
particle technique used in this work. We tested the implementation using both the Singular
Isothermal Sphere test \citep{Shu_1977} and by analysing the results from a Rotating Cloud test
\citep{Boss_1979, Burkert_1993}.

\subsection{Collapse of a Singular Isothermal Sphere}
\begin{figure} \label{IsothermalMass}
  \subfloat[The Mass of the \smartstar particle as a function of time. The slope of the
    line is approximately $1 \times 10^{-4}$ \msolar yr$^{-1}$ in excellent agreement with the
    analytical predictions \citep{Shu_1977}.]{%
  \includegraphics[clip,width=\columnwidth]{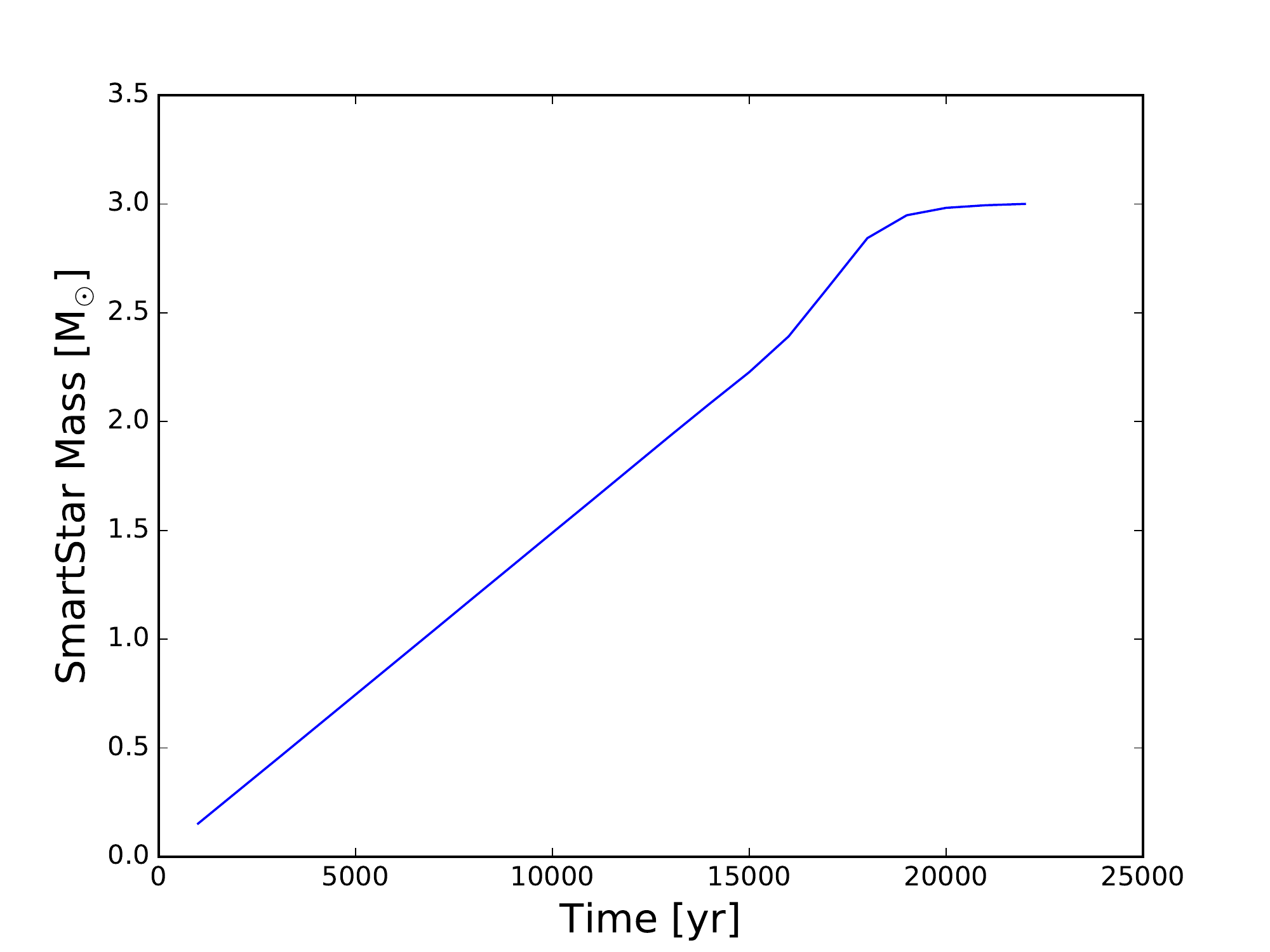}
}
  
  \subfloat[The mass accretion history of the \smartstar particle over the timeframe of the
    isothermal collapse. The mass accretion rate is constant at $1 \times 10^{-4}$ \msolar yr$^{-1}$
  until the envelope is nearly entirely accreted at T $\sim 17000$ yrs.]{%
  \includegraphics[clip,width=\columnwidth]{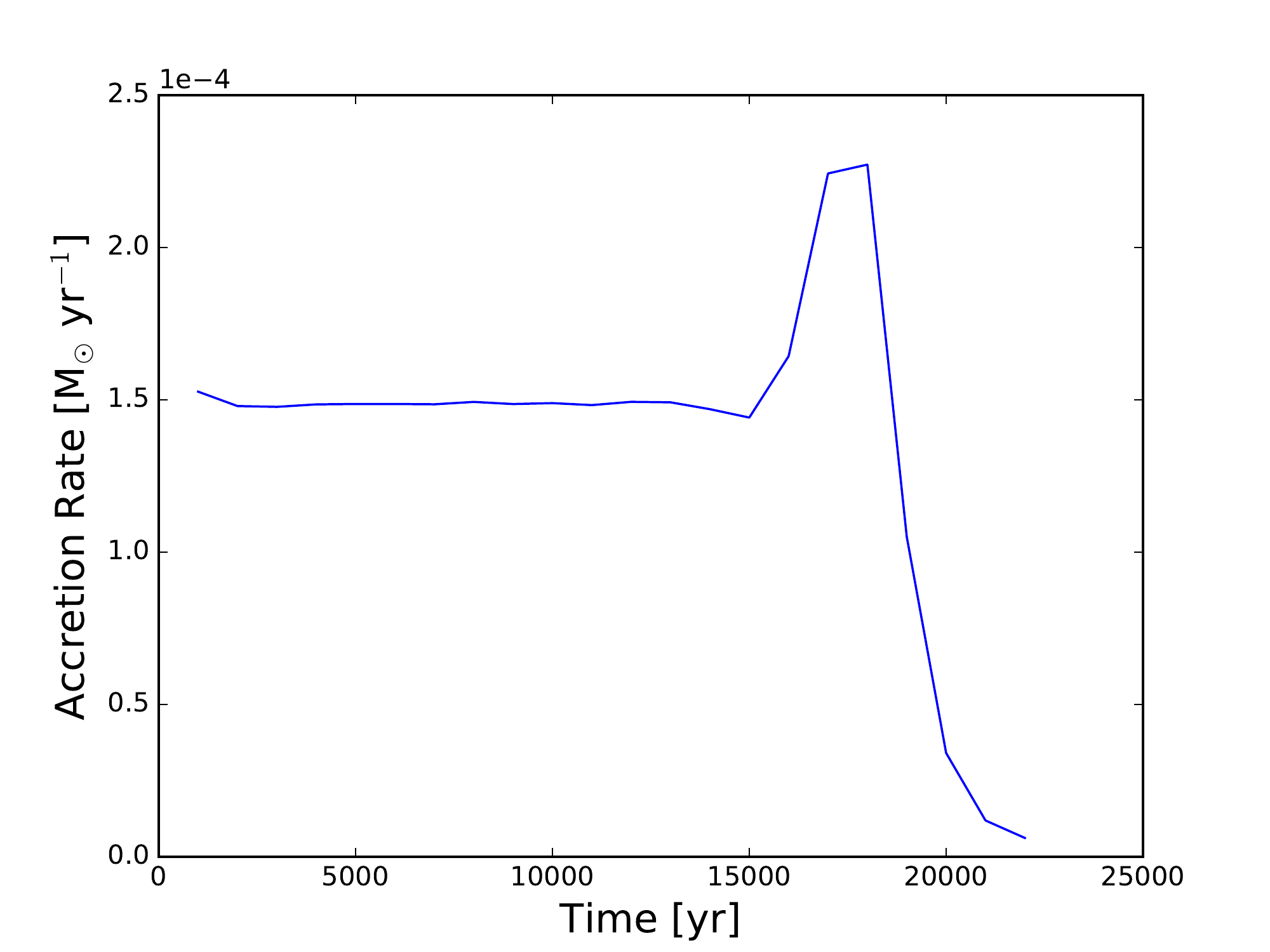}%
}

  \caption{The characteristics of the \smartstar particle undergoing isothermal collapse. In the
    top panel (a) we show the mass of the \smartstar particle as a function of time. In the bottom
    panel we show the mass accretion rate as a function of time. From both panels we see a constant
  mass accretion rate of $\dot{M} = 1.5 \times 10^{-4}$ \msolar yr$^{-1}$.}
\end{figure}

\begin{figure} 

  \subfloat[The density profile of the gas as a function of radius for multiple
    output times. The density profile is initially $\rho(R) \propto r^{-2}$ and falls to
    $\rho(R) \propto r^{-1.5}$ as the collapse proceeds. The dotted lines are the analytical
  predictions, they fall almost exactly on top of the numerical results. ]{%
  \includegraphics[clip,width=\columnwidth]{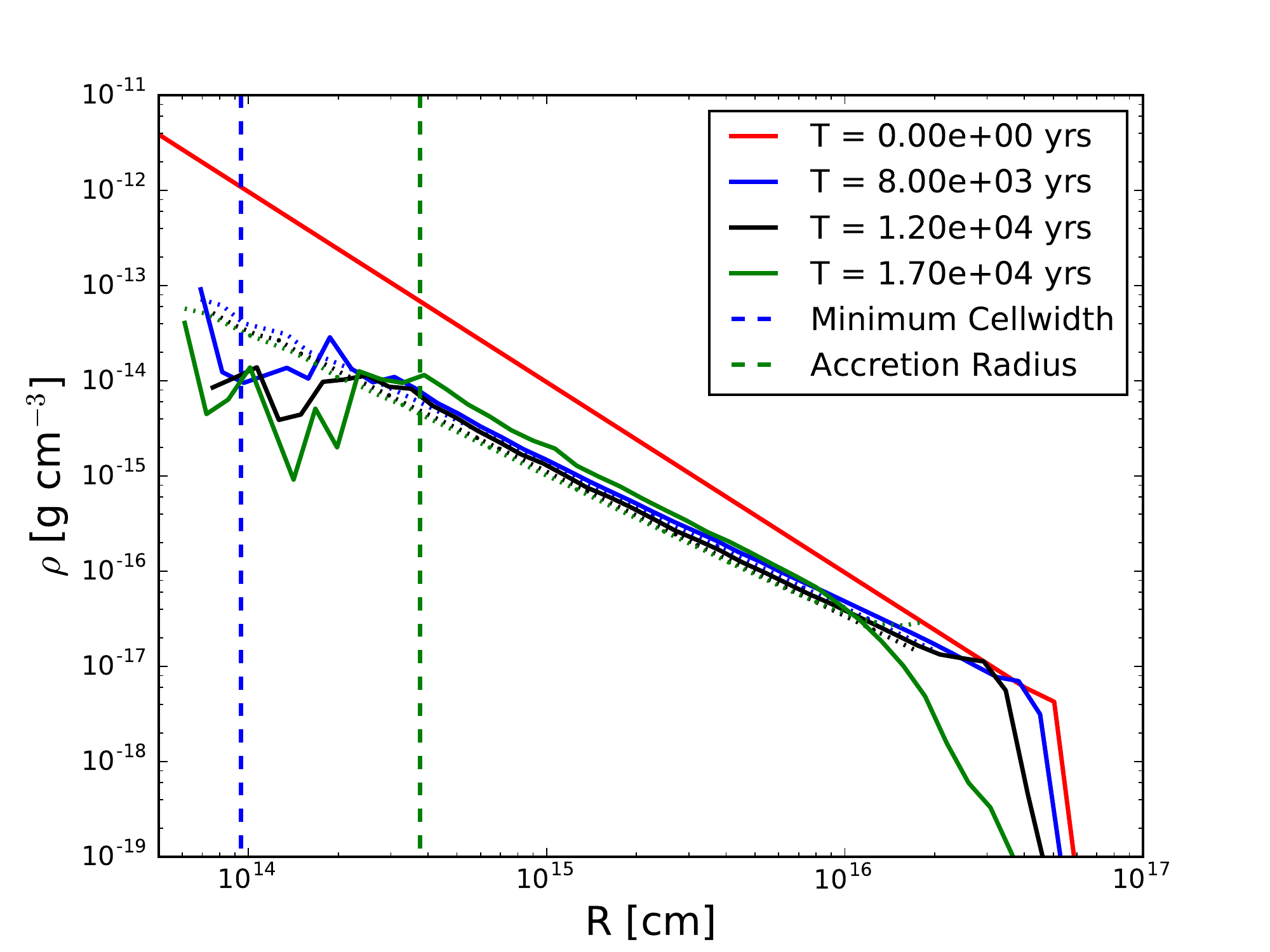}%
}

  \subfloat[The radial velocity of gas undergoing isothermal collapse. The gas is highly
  supersonic and again matches the analytical predictions (dotted lines) very closely.]{%
  \includegraphics[clip,width=\columnwidth]{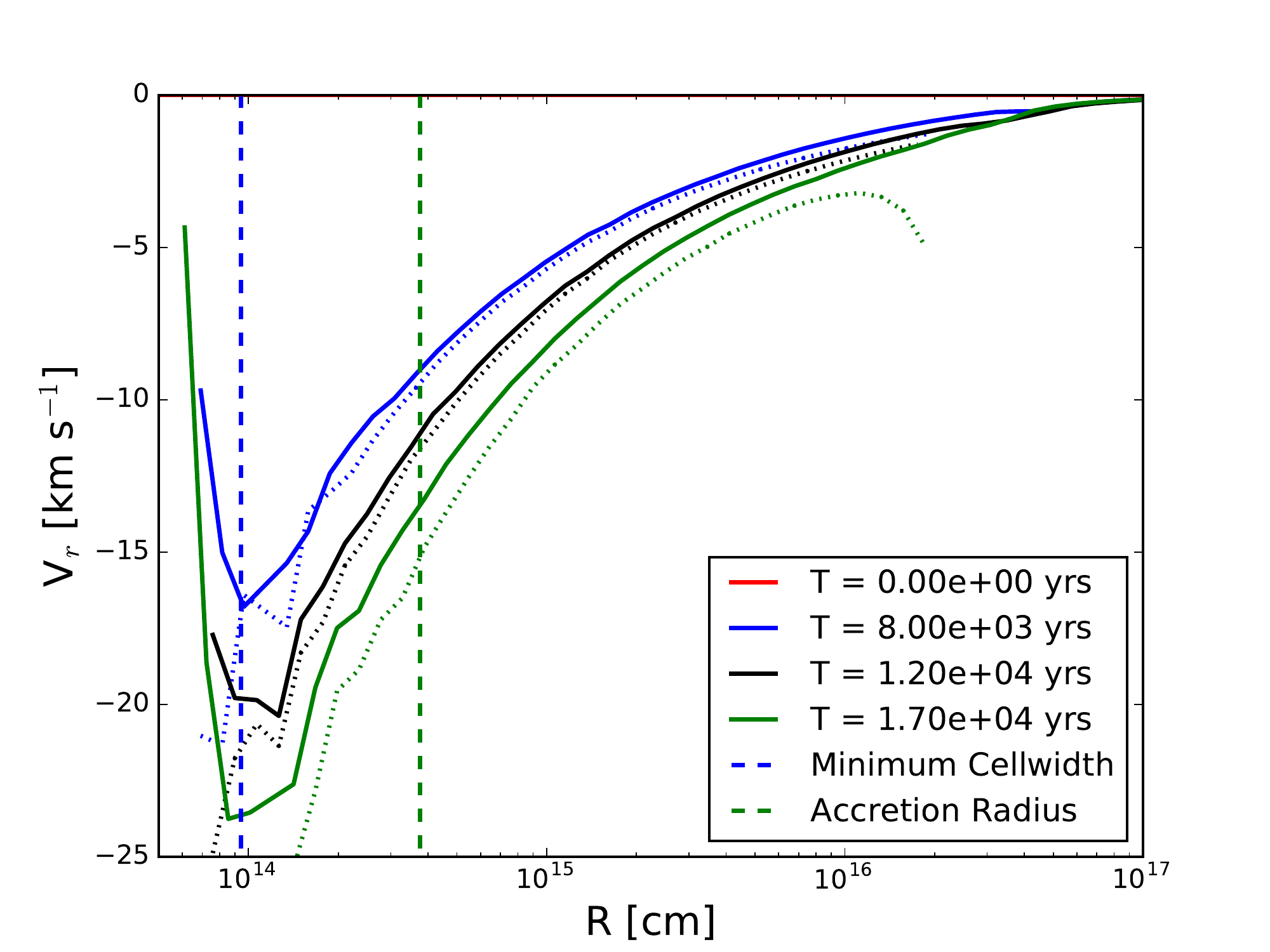}%
}

  \subfloat[The mass accretion history again showing the constant mass accretion over spherical shells
  of $\dot{M}$ $\sim 1 \times 10^{-4}$ \msolar/yr. ]{%
  \includegraphics[clip,width=\columnwidth]{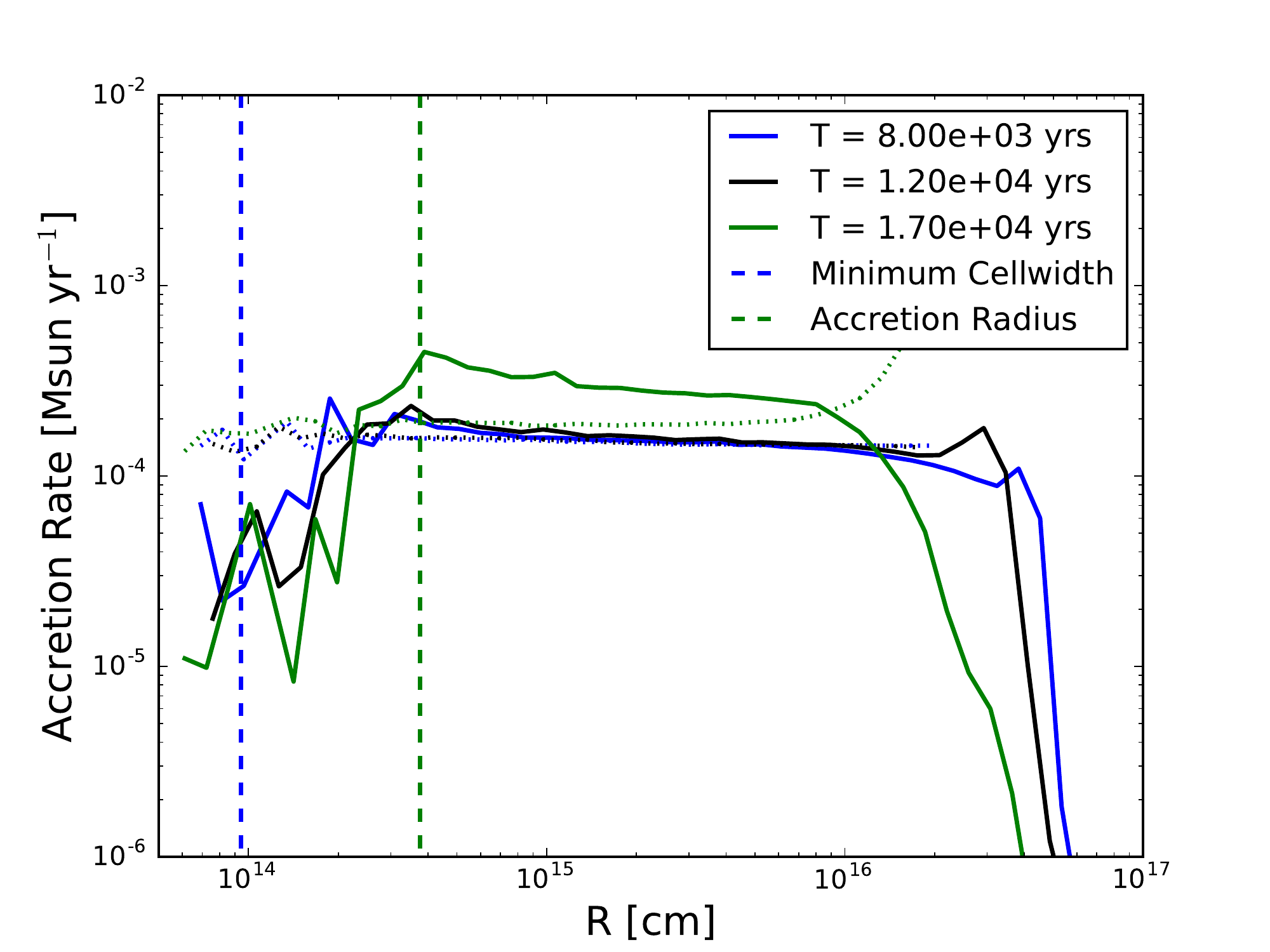}%
}

\caption{ \label{GasProfileFig} Properties of the gas surrounding the accreting \smartstar particle.}

\end{figure}

The collapse of a singular isothermal sphere with a $\rho(r) \propto r^{-2}$ density profile
produces a constant flux of mass through spherical shells i.e. a constant accretion rate.
We compute the collapse of a singular isothermal sphere to test whether our model of sink
particle accretion reflects this collapse behaviour. We follow the work of \cite{Federrath_2010} in
conducting this numerical test. The sphere is set to have a truncation radius of
R = $5 \times 10^{16}$ cm, a density at this radius of $\rm{\rho(R) = 3.82 \times 10^{-18} g\ cm^{-3}}$
and therefore a total mass of 3.02 \msolarc. The sphere is initialized at a temperature of 10 K,
the mean molecular weight of the gas is set to 3.0 resulting in a sound speed of 0.166 \kms. These
values lead to a large instability parameter for the sphere of A = 29.3 with
A = $4 \pi G \rho(R) R^2/c_s^2$ \citep{Shu_1977}. We ran the test at a maximum refinement level of
10 leading to a minimal grid spacing of $\Delta x \sim 3.05 \times 10^{-5}$ pc. No radiative cooling
or chemical reactions are calculated during the collapse. \\
\indent In the upper panel of Figure \ref{IsothermalMass} we show the mass, in solar masses,
as a function of time.
The \smartstar particle mass increases monotonically as it accretes matter eventually accreting the
entire envelope (3.02 \msolarc). The slope of the line is $1.5 \times 10^{-4}$ \msolarc/yr in
excellent agreement with the analytical solution predicted by \cite{Shu_1977}.
In the lower panel of Figure \ref{IsothermalMass} we show the mass accretion rate during this time. The test predicts
a constant accretion rate of  $1.5 \times 10^{-4}$ and this is confirmed by our test. The bump
at approximately 17000 yrs corresponds to the \smartstar particle accreting the last of the matter and
at this point the accretion rate fluctuates mildly before falling to zero. \\
\indent We also examine the behaviour of the gas itself and the mass that is accreted by the
\smartstar particle. In Figure \ref{GasProfileFig} we see the density profile, radial velocity profile
and the mass accretion through spherical shells as a function of radius. Overlaid on each plot are
the analytical expressions for the collapse taken from \cite{Shu_1977}. The agreement in all three
profiles is excellent with mild deviations only appearing once the the entire envelope has been
accreted at T $\sim$ 17000 yrs.

\subsection{Rotating Cloud Core Fragmentation Test}
\begin{figure*}
  \centering 
  \begin{minipage}{175mm}      \begin{center}
      \centerline{
        \includegraphics[width=18cm]{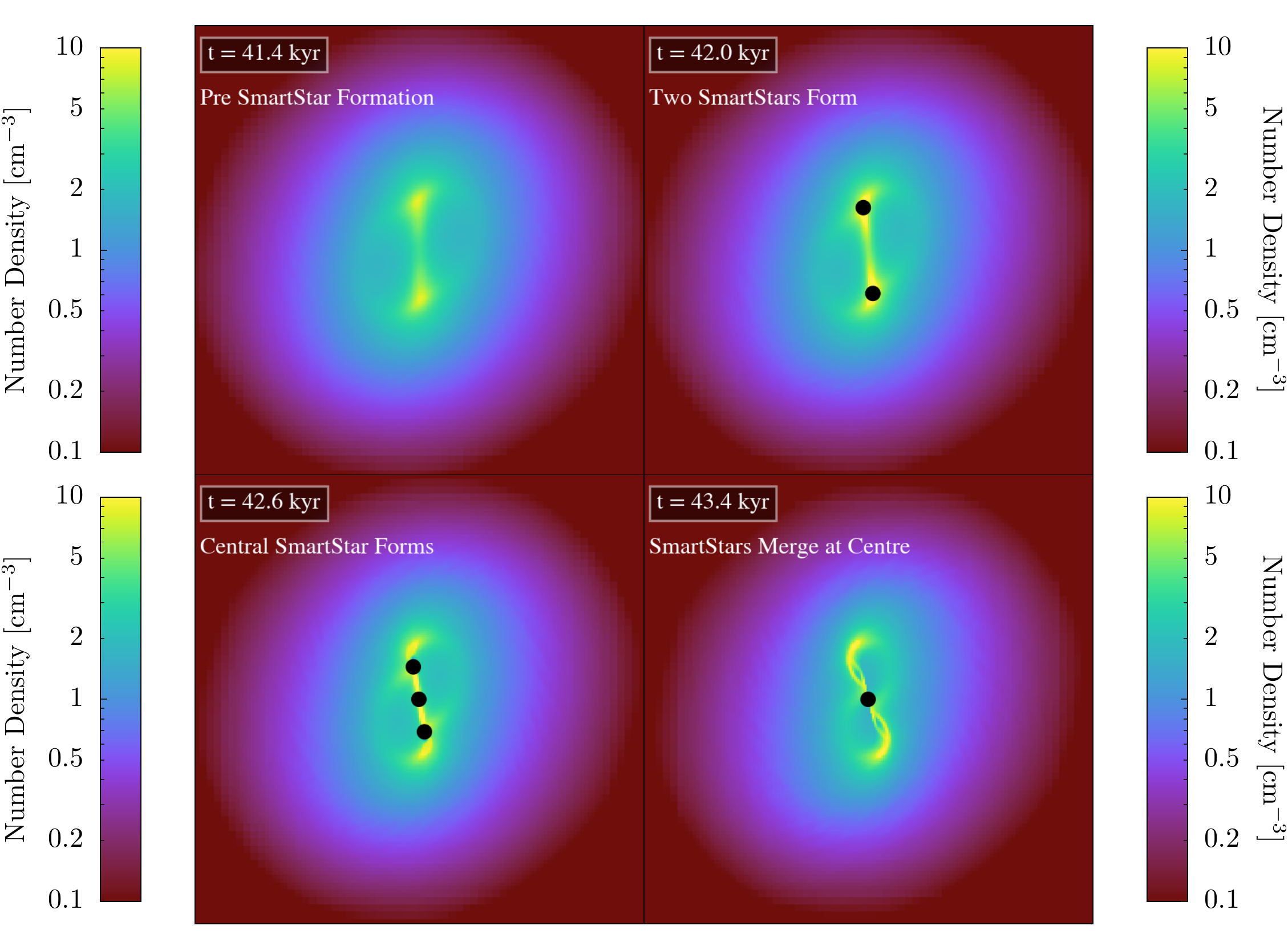}}
        \caption[]
        {\label{BurkertBodenheimer}
          Projections of the rotating cloud test uses to examine the formation
          and merging properties of the \smartstar particles. In the top left panel at t = 41.4
          kyrs the gas cloud has flattened into a disk and the threshold for \smartstar particle
          formation is about to be realised. In the top right panel \smartstar particles form
          at either end of a bar. As the simulation continues a third particle forms at the centre
          of the bar. By t = 43.4 kyrs the three \smartstar particles
          have merged at the centre of the cloud.
        }
      \end{center} \end{minipage}
  \end{figure*}

Finally, we also analyse the behaviour of the \smartstar particles when the gas is subject to
rotation and collapse. This is the classic \cite{Boss_1979} test. It is a standard test for
examining fragmentation in hydrodynamical simulations \citep{Burkert_1997, Federrath_2010}.
We employ a setup
very similar to that of \cite{Burkert_1993}, \cite{Bate_1997} and \cite{Federrath_2010}.
The cloud is initialised with the following parameters: radius R = $5 \times 10^{16}$ cm, constant
density  $\rm{\rho_0 = 3.82 \times 10^{-18} g cm^{-3}}$, mass = 1 \msolarc, angular velocity,
$\rm{\Omega = 7.2 \times 10^{-13} rad s^{-1}}$ (ratio of rotational to gravitational energy
$\beta = 0.16$), sound speed = 0.166 \kms (ratio of thermal to gravitational energy $\alpha = 0.26$),
global free-fall time $\rm{t_{ff} = 1.075 \times 10^{12} s =  3.41 \times 10^4 yr}$. The cloud is then
subject to a density perturbation with an m = 2 mode: $\rho = \rho_0[1 + 0.1\ \rm{Cos}(2\phi)]$, where
$\phi$ is the azimuthal angle. We use an isothermal equation of state throughout the test. \\
\indent The initial rotation ($\beta = 0.16$) forces the gas cloud to collapse to a disk with a
central bar due to the m = 2 initial density perturbation. \smartstar particles then form at the
ends of the bar once the conditions for particle formation are met. The test was run at a maximum
refinement level of 5 with a minimum cell size of 16.32 AU ($2.44 \times 10^{14}$ cm). 
The evolution of the system is shown in Figure \ref{BurkertBodenheimer}. The figures show the
initial formation of the disk and bar structure face on. Two \smartstar particles form at either
end of the bar at the same time due to the inherent symmetry of the system. A third fragment then
forms at the centre of the cloud core. Subsequently the two end fragments move along the
bar and merge with the central fragment. The results of this test are
in excellent agreement with similar tests carried out by \cite{Federrath_2010}.  

\label{lastpage}
\end{document}